\documentclass[12pt]{iopart}
\usepackage{color,graphicx}
\usepackage{txfonts}
\usepackage{iopams, amssymb}

\begin{document}

\article{FAST TRACK COMMUNICATIONS}{Phase diagram and strong-coupling fixed point in the disordered O($n$) loop model}

 \author{H Shimada$^1$, J L Jacobsen$^{2, 3}$, and Y Kamiya$^4$}
 \address{$^1$Mathematical and Theoretical Physics Unit, OIST Graduate University, 1919-1 Tancha, Onna-son, Okinawa 904-0495 Japan}
 \address{$^2$Laboratoire de Physique Th\'eorique de l'Ecole Normale Sup\'erieure,  24 rue Lhomond, 75231 Paris, France}
 \address{$^3$Universit\'e Pierre et Marie Curie, 4 place Jussieu, 75252 Paris, France}
 \address{$^4$Theoretical Division, T-4 and CNLS, Los Alamos National Laboratory, Los Alamos, New Mexico 87545, USA}

 \eads{\mailto{hirohiko.shimada@oist.jp}, \mailto{jacobsen@lpt.ens.fr}, \mailto{kamiya@lanl.gov}}

 \begin{abstract}
   We study the phase diagram and critical properties of the two-dimensional (2D) disordered O($n$) loop model. 
The renormalization group (RG) flow is extracted from the landscape of the effective central charge $c$ obtained by the transfer matrix method.
   We find a line of multicritical fixed points (FPs) at strong randomness for $n > n_c \sim 0.5$. We also find a line of stable random FPs for $n_c < n < 1$, whose $c$ and critical exponents agree well with the $1-n$ expansion results.
   The multicritical FP at $n=1$ has $c=0.4612(4)$, which suggests that it belongs to the universality class of the Nishimori point in the random-bond Ising model. For $n>2$, we find another critical line that connects the hard-hexagon FP in the pure model to a finite-randomness zero-temperature FP.
  \end{abstract}


\section{Introduction}

Critical properties of quenched random systems are of fundamental experimental and theoretical relevance.
2D applications include disordered magnets, polymers in random environments,
and localization in electron gases.
For instance, deriving critical exponents for the Chalker-Coddington network model~\cite{CC}, which may describe
the Anderson transition in the integer quantum Hall effect (QHE)~\cite{QHEreview}, remains an outstanding challenge
despite recent progress on a truncated model based on the supersymmetry method~\cite{Ikhlef} and exact results
for the spin QHE~\cite{Gruzberg,Bondesan}.

  Conformal field theory (CFT) provides exact critical exponents of infinite families of 2D pure models~\cite{BPZ, FQS}.
  Hence, the study of isolated FPs induced by relevant randomness in 2D deserves particular attention, because infinite-dimensional conformal symmetry is expected after disorder averaging.
  Despite much recent progress on logarithmic CFT~\cite{LCFTreview,LCFTCardy,Cardylog2,Cardylog}, applications of CFT to 2D random systems remain elusive.
  According to the Harris criterion~\cite{Harris}, pure FPs are stable against infinitesimal randomness coupled to the local energy density if the specific heat exponent, $\alpha^{}_\mathrm{pure}$, is negative.
  Previous studies confirmed that critical properties can be understood based on perturbation theory in
  the 2D random-bond Ising model (RBIM) ($\alpha^{}_\mathrm{pure} = 0$)~\cite{Maassarani} and the $q$-state random-bond Potts model (RBPM) for $2 < q < 4$ ($\alpha^{}_\mathrm{pure} > 0$)
\footnote{The first-order transition for $q>4$ is softened~\cite{JC}, confirming a general theorem~\cite{Aizenman}.}
~\cite{Ludwig,LudwigCardy,DotsenkoPP,JengLudwig,JC,JLJ}.
  However, unbiased non-perturbative studies are necessary to unveil new FPs, if any, that are not accessible by perturbation theory.
  Moreover, strong-coupling random FPs are in general not excluded \textit{a priori} even if $\alpha^{}_\mathrm{pure} < 0$.

\begin{figure}[t]
   \includegraphics[bb=0 0 402 228,height=0.25\vsize]{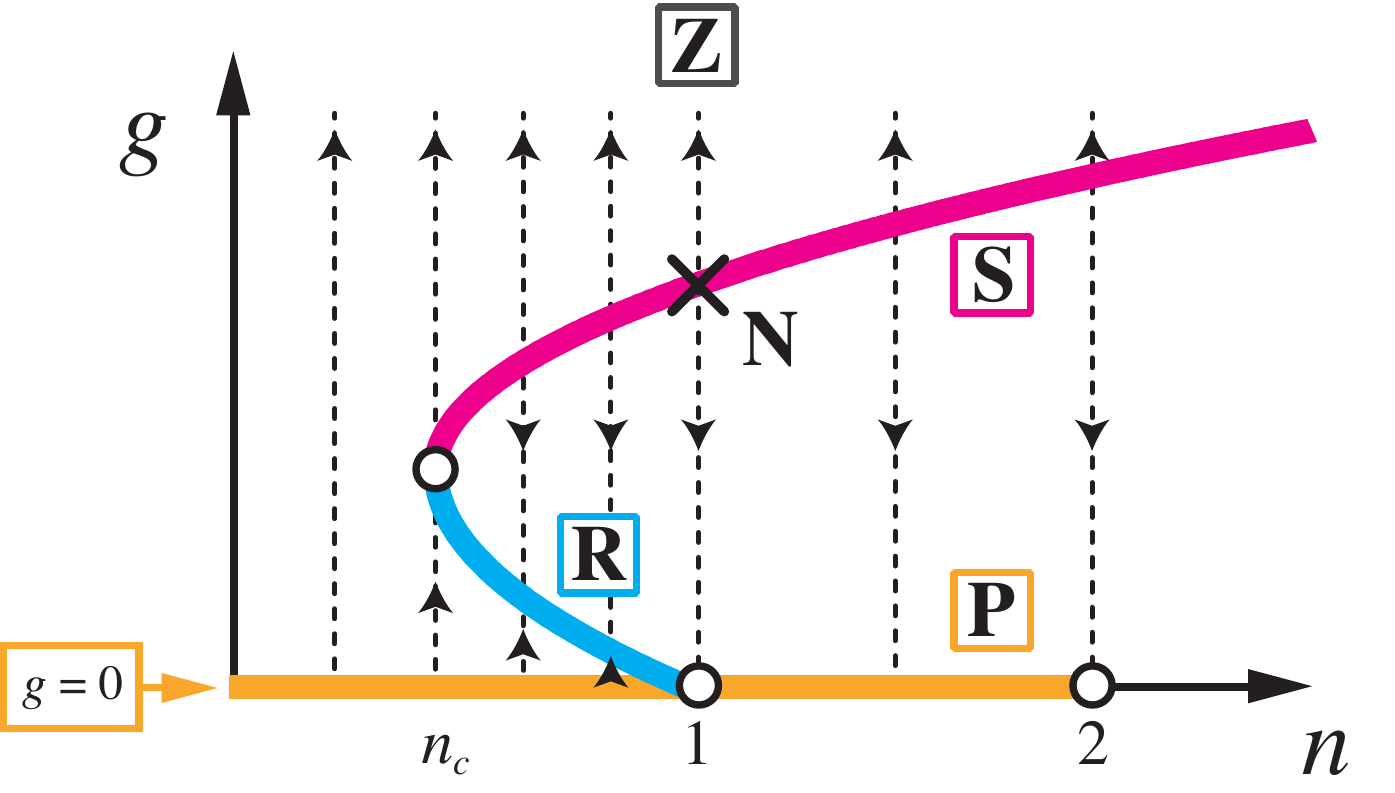}
    \caption{%
      \label{fig:RG}	
      The schematic projected RG flow along the highest temperature critical line.
      The randomness coupling, $g$, increases as disorder increases. A line of critical FPs ({\bf R}) emanates from the pure FP at $n=1$, overturning at $n=n_c$ and becomes a line of multicritical strong coupling FPs ({\bf S}). {\bf N} denotes the Nishimori point at $n=1$ (see text). 
    } 
\end{figure}

In this Communication,
we investigate the disordered O($n$) loop model, another nontrivial case where randomness is relevant (for $n < 1$~\cite{Shimada}).
The absence of self-duality ``on average'' makes its study considerably harder than the case of the RBPM, but this 
effort is rewarded by a richer phase diagram (Fig.~\ref{fig:RG}), which includes both perturbative and non-perturbative FPs.

The pure O($n$) loop model~\cite{Nienhuisloop, Jacobsenloop}
is a truncation of the lattice-regularized O($n$) non-linear sigma model~\cite{Nienhuis},
where the loop fugacity $n$ can take continuous values.
A resemblance with the $q=n^2$ tricritical Potts model extends to the continuum limit~\cite{NienhuisTricritical}, which can be analyzed by Coulomb gas~\cite{Nienhuis}, CFT~\cite{Dotsenko}, or stochastic Loewner evolution~\cite{Bauer} methods.
For $n=1$ the loops are the domain walls in the dual-lattice Ising model~\cite{Cardyloop}.  
However, important differences arise upon adding quenched bond disorder:
The energy operator $\mathcal{E}$, whose quadratic form appears in the replica approach, is identified with the primary field $\phi_{1,3}$ in the O($n$) model, 
but with $\phi_{2,1}$ in the RBPM.
Consequently, the RBPM $q-2$ expansion~\cite{Ludwig,LudwigCardy,DotsenkoPP,JengLudwig} is replaced by a $1-n$ expansion in the O($n$) model~\cite{Shimada}. 
The one-loop RG calculation suggests that a nontrivial random universality class may exist for $n_c<n<1$, with $n_c\approx 0.262$ (Fig.~\ref{fig:RG}). 
This study also suggests that the disordered polymer problem ($n \to 0$) may flow under RG to an infinite-randomness FP, where the perturbative approach is no longer reliable.

We study the disordered O($n$) loop model on the honeycomb lattice by transfer matrix (TM)~\cite{JC} and worm Monte Carlo (MC) methods~\cite{Prokofev, Wolff, Liu}.
The TM method produces the phase diagram and RG flows, while MC simulations yield the loop-size distribution and various critical exponents.
We find, in particular, a strong-randomness multicritical FP, which at $n=1$ is suggested to belong to the Nishimori universality class~\cite{Honecker, GruzbergIsing, Merz}.
Concretely, the effective central charge agrees well with that of the Nishimori point,
an unstable FP which lies at the intersection between the Nishimori line and the paramagnetic-ferromagnetic phase boundary 
connecting the pure Ising FP and the zero-temperature spin-glass point in the 2D $\pm J$ RBIM~\cite{Honecker}.
The strong-randomness FP in our model is not accessible by perturbation theory.

\section{Model and methods}
The pure O($n$) loop model is defined by $Z = \mathrm{Tr}\, \prod_{\left\langle i j \right\rangle} \left( 1+ x \bi{s}_{i}\cdot \bi{s}_{j}\right)$ with a normalization $| \bi{s}_i |^2 = n$. 
The factor per link $\left\langle i j \right\rangle$ 
is a truncation of the Boltzmann weight $\exp(x \bi{s}_{i}\cdot \bi{s}_{j})$ for the $n$-component spin model~\cite{Nienhuis}.
Interestingly, the loop model correctly describe the universality class of the critical spin model, since the omitted terms like 
$(\bi{s}_{i}\cdot \bi{s}_{j})^2$ are RG irrelevant \cite{Nienhuisloop}.
Because of the global O($n$) symmetry, integrating out spins on the honeycomb lattice yields
\begin{equation}
  Z =  \sum_{\mathrm{loops}}  x^\mathcal{B}  n^\mathcal{L},
  \label{looppartition}
\end{equation}
where the sum runs over all possible loop configurations, each with $\mathcal{B}$ 
occupied bonds and  $\mathcal{L}$ loops.
Here, the fugacity $n$ is naturally extended to generic non-integer $n$ and the region with a positive statistical weight is enlarged from $ | x | < 1/n$ in the spin model to any $x > 0$.
The phase diagram of the O($n$) honeycomb lattice loop model is well understood.
The model undergoes a continuous phase transition for $|n|\leq 2$ in the same universality class as the spin model.
Three-state Potts universality is observed for $n>2$ in the large-$x$ regime specific to the loop model~\cite{Guo}.

The disordered model has a quenched random variable $x_{ij}$ at each link drawn from a distribution $P(x)$ instead of $x$ in \eref{looppartition}.
In our numerics we use the bimodal distribution peaked at positive $x$,
\begin{equation}
 P(x) = p \delta(x-x_1) + (1-p) \delta(x-x_2),
\label{binarydist}
\end{equation} 
parametrized by
\begin{equation}
 t= \gamma \left[ p x_1 + (1-p) x_2 \right]^{-1},~~ s = \sqrt{x_2/x_1} \in [0, 1],
\label{t_definition}
\end{equation} 
specifying the averaged temperature and randomness-strength, respectively.
We use $p=1/2$ and specify the constant $\gamma$ later.
This simple bond-randomness leads to a rich phase diagram with several nontrivial FPs.

We mainly use the flow analysis~\cite{JC, JP} that assumes existence of a potential
function $C$ for the RG flow based on the $C$-theorem~\cite{Zam}, similar to~\cite{Hukushima,Kamiya}.
We identify $C$ with the effective central charge $c$ measured through the finite-size scaling (FSS) of the averaged free energy obtained by the TM method, and find FPs as extremums of $C$. The $C$-theorem invariably predicts downhill RG flow in the landscape of $C$ in unitary systems \cite{Zam}. On the contrary, the non-unitarity of our random model allows for both downhill and uphill flows; the non-standard uphill flows could be understood from a sign change of Zamolodchikov's metric in the replica limit \cite{Shimada} and seems common in random systems. For instance, uphill flow along the randomness was predicted \cite{LudwigCardy} and confirmed numerically \cite{JC, JP, Picco} for the RBPM with $q > 2$. In the vicinity of the pure FPs, the relation between the landscape gradient and the RG flow can be determined numerically with the aid of the Harris criterion \cite{Harris}. As we describe below, in most cases, the direction of the flow for weak randomness is uphill along the randomness while it is downhill along the temperature.\footnote{
Here ``most cases" refers to finite temperature RG, which is our focus. Special care might be needed for $t=0$.}
In contrast to the RBPM, one cannot exactly determine the critical averaged temperature of the random FP because of the absence of self-duality
(corresponding to the appearance of $\mathcal{E} \sim \phi_{1,3}$ on the right-hand side of 
the operator product expansion $\mathcal{E}\cdot\mathcal{E}\sim I+\mathcal{E}+\mathcal{E}'$, except for $n = 1$).
Thus we need to search possible FPs on the two-parameter theory space $(t ,s)$. 

The TM is constructed over the connectivity basis on a
strip of width $L$ with periodic boundary conditions~\cite{Blote}. Its dimension is the Motzkin number $M_L$,
the number of ways of drawing any number of non-intersecting chords joining $L$ points on a circle. 
For the honeycomb lattice, CFT predicts the FSS form of the free energy per site as~\cite{BloteCardyNightingale,Affleck}
\begin{equation}
  f(L)= f(\infty) - \frac{A \pi c}{6 L^2}+ \frac{d}{L^4}+\cdots \,,
  \label{fss_CFT}
\end{equation}
with the effective central charge $c$, a non-universal coefficient $d$~\cite{JC}, 
and the constant $A = 2/\sqrt{3}$ being the density of sites per unit area.

Data collection involves typically $10^3$ independent realizations of strips of length $10^7$.  
Error bars are estimated from the standard deviation of $f(L)$ on patches of various
lengths $\sim 10^3$. As in~\cite{JC}, we subject $f(L)$ to 
two- or three-point fits using $c$ and $d$ as fitting parameters for various widths $L=L_0$, $L_0+2$ (and $L_0+4$). 
We choose $L_0 \sim 6$ so that the errors due to finiteness of $L_0$ and randomness are comparable.

\section{RG landscape} \label{RG}
To trace out the landscape of the $C$-function, 
we start from the pure section $s=1$ with four exact reference points for $|n| \leq 2$. 
First, the pure O($n$) loop model has a critical FP ({\bf P}) with central charge
$c_\mathrm{P}^{}=1-6(1-\rho)^2 / \rho$, with $\rho\in [\frac{1}{2},1]$ and $n=-2\cos (\pi/\rho)$.
\footnote{This parameter is the squared negative-charge $\alpha_-^2$ (e.g., $\rho=\frac34$ for the Ising model) in CFT.}
Second, the low-$T$ critical ``dense'' phase~\cite{Nienhuisloop} corresponds to another FP ({\bf D}) with $c_\mathrm{D}^{} = 1-6(1-\rho)^2 / [\rho(2\rho-1)]$. The location of ${\bf P}$ (${\bf D}$) for the honeycomb lattice model is $x = x_+$ ($x_-$) with $x_\pm(n) = \left( 2 \pm \sqrt{2-n} \right)^{-1/2}$~\cite{Nienhuisloop}.
Third, the loop model at $x \to \infty$ acquires an enhanced symmetry due to the prohibition of empty vertices, realizing the FP ({\bf F}) of the fully-packed loop model with 
$c_\mathrm{F}^{}=c_\mathrm{D}^{} + 1$~\cite{Kondev}.  
Fourth, we have the high-$T$ trivial FP (${\bf X}$) at $x = 0$ with $c = 0$.
These FPs lie at $t_\mathrm{F}=0$, $t_\mathrm{D}=x_+(n)/x_-(n)$, $t_\mathrm{P}=1$, and $t_\mathrm{X}=\infty$ with $\gamma=x_+(n)$ in \eref{t_definition}, which is assumed hereafter unless otherwise mentioned.
As $t$ increases along $s=1$ from $t = 0$, $c$ has a minimum at {\bf D} and a maximum at {\bf P}, namely the RG flow in the pure system is downhill ({\bf F} $\Rightarrow$  {\bf D} $\Leftarrow$ {\bf P} $\Rightarrow$ ${\bf X}$), in agreement with the $C$-theorem for unitary systems~\cite{Zam}.

\begin{figure}[t]
  \includegraphics[bb=0 0 551 292, width=0.8\hsize]{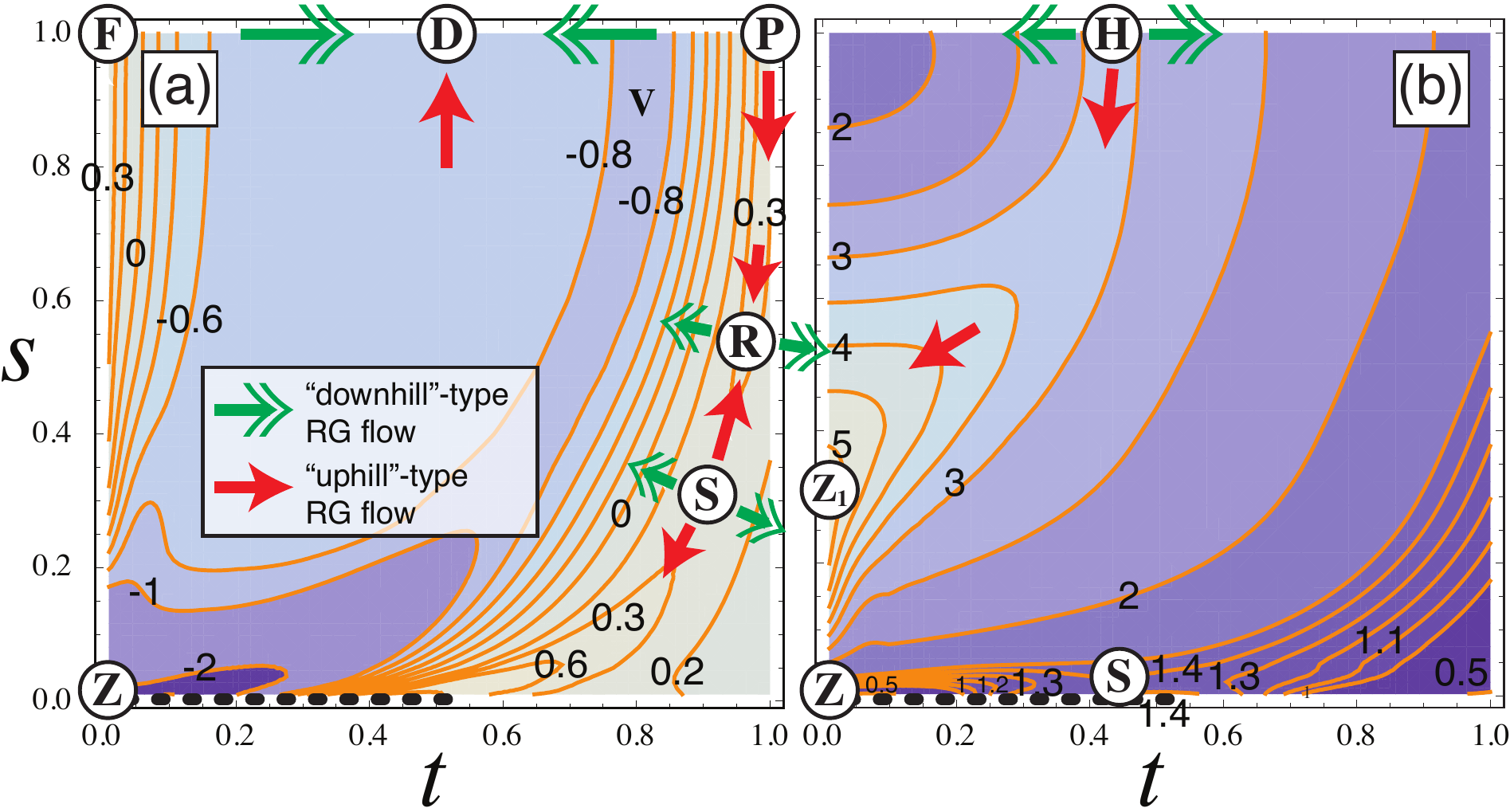}
    \caption{%
      \label{fig:C}
      (a) The contour plot (``landscape'') of $C(t,s)$ for $n=0.6$ ($\gamma = x_+(0.6)$). The random ({\bf R}), strong-randomness multicritical ({\bf S}), and infinite-randomness ({\bf Z}) FPs lie along the ``ridge'' emanating from {\bf P}. The valley {\bf V} is most likely unphysical (see text). The dashed line indicates almost degenerated contours near {\bf Z}. Arrows represent RG flows near FPs.
      (b) $C(t,s)$ for $n=8$ ($\gamma = 1/\sqrt{2}$). The hard hexagon FP ({\bf H}) flows into the zero-temperature FP (${\bf Z_1}$). 
    } 
\end{figure}

Next we switch on the randomness. 
The landscape $C(t, s)$ is plotted with contours for $n=0.6$ in Fig.~\ref{fig:C}(a).
The whole surface of $C(t, s)$ is a simple homotopic deformation of the curve of $C$ 
from $s=1$ (pure) to $s=0$ (infinite randomness) except for the vicinity of the infinite-randomness FP ({\bf Z}) at $(t, s)=(0, 0)$. We clarify universality classes of {\bf R} and {\bf S} along the critical line {\bf PRSZ}, 
while the detailed study along the zero-temperature $t=0$, and especially the nature of {\bf Z} is beyond the scope of this article.
The valley {\bf V} between {\bf D} and {\bf P} in $ | {1 - s} | \ll 1$, where the pure model has no FPs, is most likely a numerical artifact as it becomes narrower as $L_0$ increases.
Most of the low but finite $t$ regime flows into {\bf D}.
The above qualitative feature changes at $n=1$, where {\bf R} is absorbed into {\bf P}.
The other qualitative change occurs at $n=n_c$ with
\begin{equation}
  0.5 \lesssim n_c < 0.55,
  \label{n_c}
\end{equation}
where pair annihilation of {\bf R} and {\bf S} takes place (see below).

\subsection{Random fixed point}
We now turn to the RG flow along the critical line {\bf PRSZ} 
[``ridge'' in Fig.~\ref{fig:C}(a)]
for $n_c < n < 1$.
Our numerics show that {\bf P} becomes a saddle point (a peak) for $n<1$ ($n>1$). 
Since the Harris criterion~\cite{Harris} tells us that the flow should deviate from (return to) {\bf P} in its vicinity for $n<1$ ($n>1$),
we conclude that the RG flow along the critical line corresponds to the uphill direction, consistent with the non-unitary variant of the $C$-theorem~\cite{JC, JP, Shimada}.

\begin{figure}[t]
    \includegraphics[bb=0 0 370 209,width=0.7\hsize]{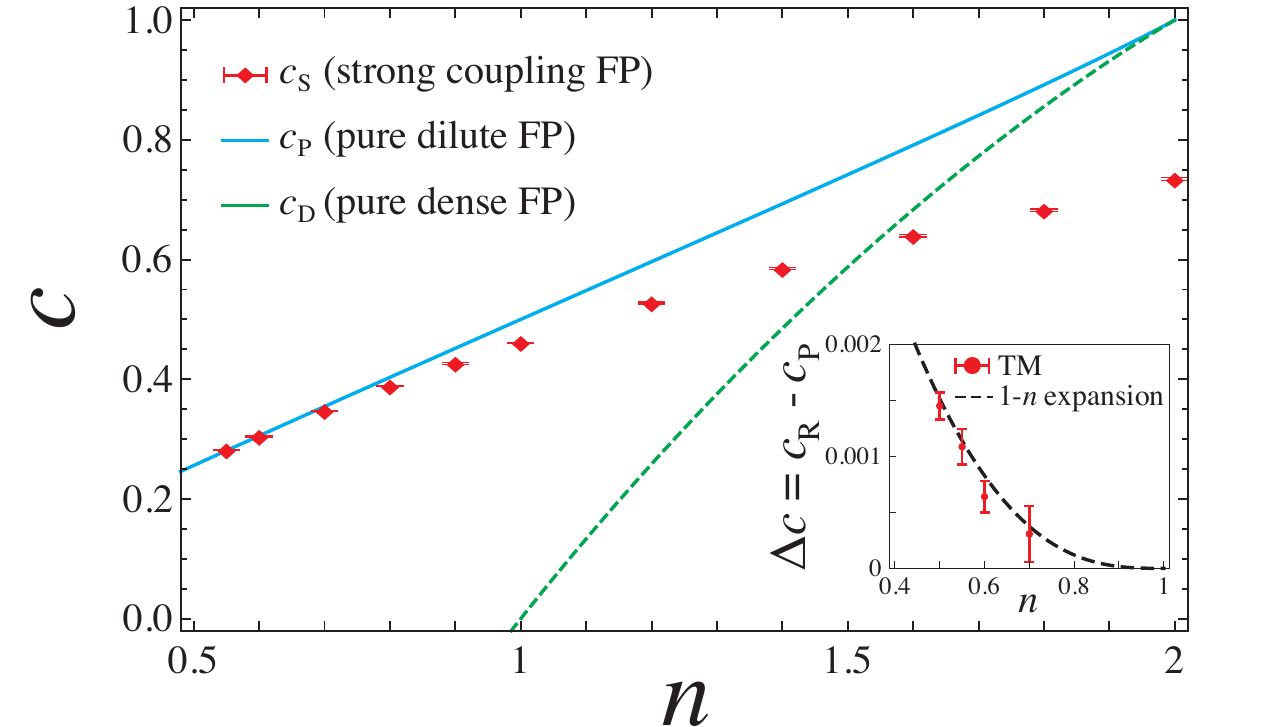}
    \caption{%
      \label{fig:global_c}
      The central charge $c^{}_\mathrm{S}$ of {\bf S} as a function of $n$. 
      $c^{}_\mathrm{P}$ and $c^{}_\mathrm{D}$ are shown for comparison. $c^{}_\mathrm{R}$ nearly coincides with $c^{}_\mathrm{P}$ in its whole regime $n_c<n<1$ but the nontrivial shift, $\Delta c=c_\mathrm{R}^{}-c_\mathrm{P}^{}$ (see the inset), is confirmed in a good agreement with the prediction of \cite{Shimada}.
    } 
\end{figure}

We first discuss the case with $n$ slightly below $1$.
The ridge in Fig.~\ref{fig:C}(a) has two extremums, a peak {\bf R} and a saddle point {\bf S}, apart from {\bf P}.
We identify {\bf R} and {\bf S} as, respectively, the stable random FP found in the $(1-n)$ expansion~\cite{Shimada}, and a novel multicritical point. We plot the corresponding central charges $c_\mathrm{R}^{}$, $c_\mathrm{S}^{}$ in Fig.~\ref{fig:global_c}.
Although the shift $\Delta c=c_\mathrm{R}^{} - c_\mathrm{P}^{}$ is small, the numerical precision in the region $0.5 \lesssim n \lesssim 0.7$ suffices to establish a good agreement with the one-loop calculation on {\bf R} (see the inset).

More importantly, our TM calculation establishes the absence of random FPs in the finite-randomness regime for $n < n_c$
due to the pair annihilation of {\bf R} and {\bf S} (see Fig.~\ref{fig:RG}).
At $n=0.55$, {\bf R} is found at $(t,s) \approx (0.94608, 0.4504)$ with  $c_\mathrm{R}^{}=0.28182(16)$.
At the same $n$, {\bf S} is found at $(t,s) \approx (0.91828, 0.3523)$ with $c_\mathrm{S}^{} = 0.2817(2)$, which is very close to $c_\mathrm{R}^{}$.
On the contrary, we observe a monotonic flow from {\bf P} to {\bf Z} along the critical line for $n < n_c$.
In particular, this suggests that the disordered polymer may be governed by {\bf Z}.
The discrepancy of the critical value in \eref{n_c} from the one-loop estimation $n_c\approx 0.262$~\cite{Shimada} is not surprising.
To obtain analytically a better estimate of $n_c$, one should at least include the $g^3$ term in the beta function ($g$ is the randomness coupling in field theory).
At present, this term has been calculated only in an approximation that assumes the structure constant $C^{\mathcal{E}}_{\mathcal{E}\mathcal{E}}$ is $\mathcal{O}(1-n)$~\cite{Shimada}, which may not be justified at $n\sim n_c$.

\begin{figure}[bt]
  \includegraphics[bb=0 0 357 188,width=0.7\hsize]{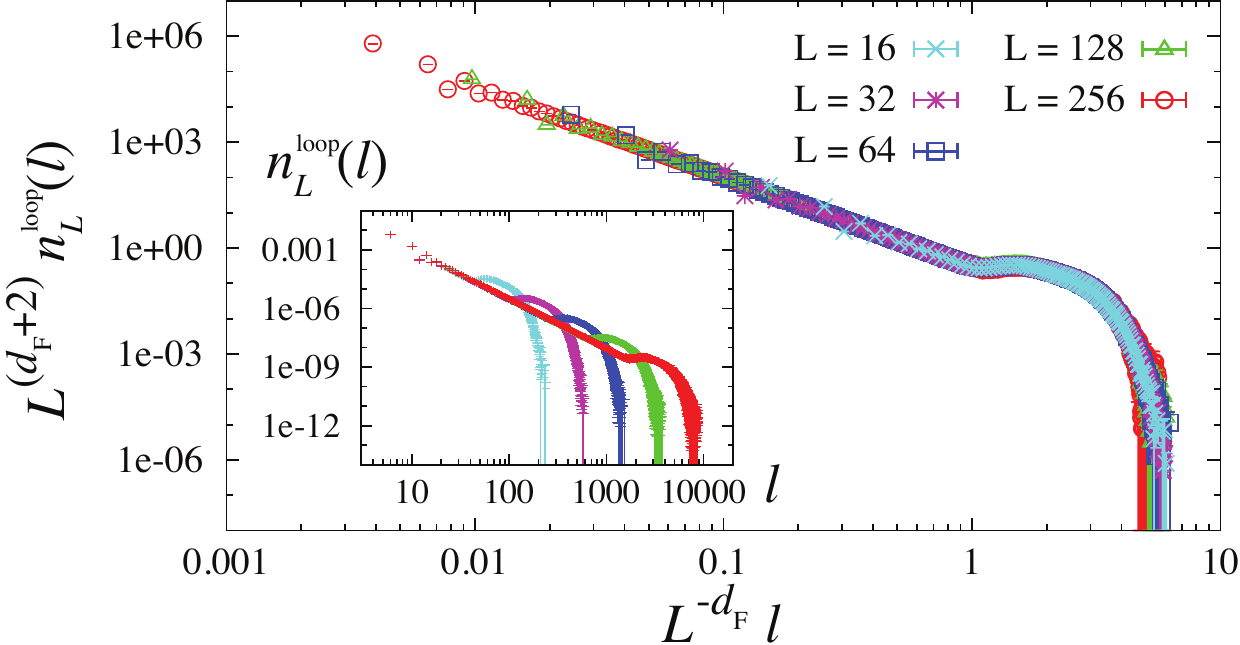}
    \caption{%
      \label{fig:loop}
      The FSS of the loop-size distribution $n_{L}^\mathrm{loop}(l) \sim L^{-(d_\mathrm{F}+2)} \Phi(L^{-d_\mathrm{F}}l)$ for $n = 0.55$ at $(t,s)=(0.946046,0.450444)$ ({\bf R}) with $\gamma = x_+(0.55)$ measured on the $L \times L$ honeycomb lattice by using the MC simulation. The inset shows $L$-dependence of the estimate of $d_\mathrm{F}$ from the three-point fit $(L, 2L, 4L)$ of designated quantities. 
      The dashed line indicates the pure FP value.
    }
\end{figure}

We determine the scaling dimension $x_\sigma$ of the spin operator at {\bf R} by computing 
the probability $G(r)$ that one string propagates a distance $r$ along the cylinder. 
This probability for a given sample is related to the free-energy gap between the one- and zero-string sectors: $\Delta f_1(L) = f_1(L)-f(L) =-\frac{1}{Lr}  \ln G(r)$. The cumulant expansion~\cite{JC} enables one to express the disorder average in terms of the self-averaging cumulants:
$\ln \overline{G} =  \overline{\ln G} + \frac12  \overline{(\ln G -\overline{\ln G})^2} + \cdots,$ 
and then to compare this with the CFT prediction $-\frac{1}{Lr}\ln \overline{G} = \frac{2 A \pi  x_\sigma}{L^2} + o(L^{-2})$~\cite{Cardy84}.
We used the first three cumulants calculated from $10^9$ cylinders of length $r=100$. The measured value $x_{\sigma, \mathrm{R}}=0.1174(4)$ for $n=0.55$ is consistent with the prediction of the $1-n$ expansion $x_{\sigma, \mathrm{two-loop}} \approx 0.11748$~\cite{Shimada} and larger than $x_{\sigma, \mathrm{P}} \approx 0.11647$ for the pure model.
Meanwhile, the worm MC simulation gives $x_{\sigma, \mathrm{R}}=0.1185(15)$, which is again larger than $x_{\sigma, \mathrm{P}}$ and in reasonable agreement with $x_{\sigma, \mathrm{two-loop}}$.
We also estimate the loop fractal dimension $d_\mathrm{F}$ at {\bf R} characterizing the typical loop size relative to the correlation length, $l^\ast \sim \xi^{d_\mathrm{F}}$~\cite{Janke}, by measuring the loop-size distribution in the MC simulation (Fig.~\ref{fig:loop}). 
We obtain $d_\mathrm{F,\mathrm{R}} = 1.32(1)$ for $n=0.55$, slightly smaller than the one at the pure FP, $d_\mathrm{F,\mathrm{P}} \approx 1.35428$~\cite{Nienhuisloop, Jacobsenloop}.

\subsection{Strong-coupling multicritical fixed point}
At $n=1$, the saddle point {\bf S} is located at $(t,s) = [0.81585(10), 0.1904(2)]$ with 
\begin{equation}
  c_\mathrm{S}^{} = 0.4612 \pm 0.0004.
  \label{nishimori_central}
\end{equation}
This is clearly distinct from $c=\frac{5\sqrt{3}\ln 2}{4\pi} \approx 0.477$ of percolation on Ising clusters~\cite{JC}.
Instead it lies well inside the error bar of the result $c=0.464(4)$~\cite{Honecker} found for the Nishimori FP in the 2D $\pm J$ RBIM (Edwards-Anderson model for spin glasses).
We note that the honeycomb disordered O(1) loop model is an exact (i.e., domain-wall) representation of  the RBIM on the triangular lattice, in which the dimensionless exchange coupling can take two different values $K_i = (\ln x_i) / 2$ ($i = 1, 2$) with probability $p$ and $1 - p$. Interestingly, upon this duality mapping, {\bf S} corresponds to a regime where the system has both randomness and frustration.
Note however that since $K_2/K_1 \approx -9.64 \neq -1$, this model does not have an obvious gauge symmetry.
On the other hand, the Nishimori FP in the $\pm J$ RBIM 
is located at the intersection of the Nishimori line (with local $\mathbb{Z}_2$ gauge symmetry) and the paramagnetic-ferromagnetic phase boundary~\cite{Nishimori}.
While our model does not possess such a local gauge symmetry explicitly, this does not necessarily preclude the Nishimori \textit{universality class}.
Thus, we interpret~\eref{nishimori_central} as a strong indication of this universality class at ${\bf S}$ for $n=1$.
To corroborate this, it will be interesting to verify other fingerprints of the Nishimori universality class, such as
the pairwise degeneracy of the multifractal exponents that follows either from the gauge symmetry~\cite{Nishimori, LeDoussal} or $\mathrm{Osp}(2m+1|2m)$ supersymmetry~\cite{GruzbergIsing};
this will be reported elsewhere.
Accepting this identification, our strong randomness FP {\bf S} can be seen as a one-parameter generalization
of the Nishimori point for continuous values of $n>n_c$ (see Figs.~\ref{fig:RG} and \ref{fig:global_c}).

The physics at {\bf S} is non-perturbative. This observation can be illustrated by the following beta function of the disordered $O(1)$ model (RBIM) corresponding to that of the 2D $M$-color Gross-Neveu model in the $M\to 0$ limit~\cite{Ludwig,Gracey}:
\begin{equation}
\beta  (g) = (M-2) \left[ g^2  - g^3 -\frac{(M-7)}{4} g^4\right]+ \mathcal{O}(g^5) \,.
\label{beta_g} 
\end{equation}
Here, although a fictitious FP ($c=3/8$) is obtained at $g=1$ if we truncate \eref{beta_g} at the third order, a Borel-Pad\'{e} analysis of \eref{beta_g}
reveals that a nontrivial FP at $g=\mathcal{O}(1)$ does not exist for $M \lesssim M_c=11/4$
\footnote{The situation is analogous to the $\phi^4$ theory in $d=4$, where the Borel-Pad\'e resummation correctly preclude infrared unstable FPs (unphysical roots of the truncated beta function).}.
Thus, nontrivial random FPs cannot be obtained at this level of perturbation theory.

\subsection{Structure for $n > 2$}
The phase diagram changes qualitatively at $n = 2$, where {\bf P} and {\bf D} merge and annihilate \cite{Guo}.
While the saddle point {\bf S} seems to remain also for $n > 2$, the critical line from {\bf S} to {\bf P} disappears.
Meanwhile, a new critical line emanates from the hard-hexagon (three-state Potts) FP {\bf H} in the pure limit [see Fig.~\ref{fig:C}(b)]. 
{\bf H} describes a transition where loops crystallize into hexagons (the shortest possible loops), breaking $\mathbb{Z}_3$ lattice symmetry~\cite{Guo}.
For infinite $n$, the bond randomness amounts to adding an inhomogeneous weight for each hexagon, which may 
act like a random field for the three-state Potts spins. 
Despite this analogy and the absence of long-range order in the 2D random field Potts model~\cite{Aharony},
the landscape for $n=8$ seems to show that the critical temperature decreases rather slowly as the randomness increases.
Here, the critical line from {\bf H} terminates at another zero-temperature fixed point ${\bf Z_1}$ with a finite randomness $s\sim 0.3$ and $c\sim 5$.

\section{Conclusion and outlook} \label{Conclusion}
By studying the landscape of the effective central charge, we have investigated the phase diagram for the disordered O($n$) loop model for a wide range of positive $n$.
Besides elucidating the nature of the random FP {\bf R} for $n_c < n < 1$ (see Fig.~\ref{fig:RG}), we found a strong randomness FP {\bf S} that appears for $n>n_c$, which seems non-perturbative. In particular, {\bf S} at $n=1$ is suggested to be in the Nishimori universality class known for the 2D RBIM, which could be explained based on the duality mapping to the triangular lattice RBIM, where {\bf S} indeed corresponds to a regime with randomness and frustration. 
Thus, the line of {\bf S} may be a one-parameter ($n$) generalization of the Nishimori FP.
To obtain a non-perturbative description it would be interesting to look for symmetries of {\bf S} for generic $n$ that may generalize the supersymmetry present at the Nishimori point in the $\pm J$ RBIM.

In the 2D Ising spin glass model, the Nishimori FP separates the transition line to the ferromagnetic ordered phase into one in the weak-disorder regime controlled by the pure Ising FP and the other in the strong-disorder regime, 
which is controlled by the zero-temperature FP separating ferromagnetic and spin glass phases.
It is also interesting to note that the zero-zerotemperature property might be related to the spin glass phase of the triangular lattice RBIM \cite{Poulter}.

While the nature of ${\bf Z}$ in our model is unclear, 
a configuration in the infinite-disorder limit has maximal weight if it covers all the strong bonds and none of the weak bonds.
For the RBPM this provides a mapping to the (replica limit of a) bond-percolation problem and critical exponents follow~\cite{JC}.
However, for the disordered O($n$) model there is a competition with the geometrical constraint that the covered bonds have
to form closed loops. We therefore believe that the properties of {\bf Z} could be studied using combinatorial optimization
techniques. It is likely that the parameter $p$ in (\ref{binarydist}) could play a nontrivial role, since the bond-percolation threshold
on the honeycomb lattice, $p_{\rm c} = 1 - 2 \sin(\pi/18) \simeq 0.652$~\cite{Sykes}, is larger than $p=1/2$ used in our simulations.
We observe that near ${\bf Z}$, the FSS form for first-order transitions~\cite{BloteNightingale} seems more appropriate than the CFT form \eref{fss_CFT}.

To further understand the other zero-temperature FP ${\bf Z_1}$ for $n > 2$,
it would be interesting to assess whether
the central charge $c^{}_{\mathrm{Z}_1}$ increases asymptotically as $\propto \log(n)$, as was observed for the RBPM~\cite{JC,Picco}.
It would also be interesting to study asymptotically the equivalence of ${\bf Z_1}$ with the random-field model, following~\cite{JC}.

\ack
The authors wish to thank \'Edouard Br\'ezin, Ilya A. Gruzberg, Hubert Saleur, and Cristian Batista for valuable discussions.
The work of HS was supported by a Bourse du Gouvernement Fran\c{c}ais and the Philippe Meyer foundation.
The work of JLJ was supported by the Institut Universitaire de France and the Agence Nationale de la Recherche (grant ANR-10-BLAN-0414).
The work of YK at LANL was performed under the auspices of the U.S. DOE Contract No.~DE-AC52-06NA25396 through the LDRD program.
Part of the numerical work was done on the CURIE supercomputer (GENCI-TGCC 2013-056956) as well as on the supercomputers of NERSC.

\maketitle
\end{document}